\documentclass[aps,prl,reprint,longbibliography]{revtex4-2}
\usepackage{amsmath}
\usepackage{enumitem}
\usepackage{amsthm}
\usepackage{amssymb}
\usepackage{mathtools}
\usepackage[linesnumbered,ruled]{algorithm2e}
\SetEndCharOfAlgoLine{}
\usepackage{svg}
\usepackage{subfloat}
\usepackage{placeins}
\usepackage[caption=false]{subfig}
\usepackage{multirow}
\usepackage{wasysym}
\usepackage{hyperref}
\usepackage{bm}
\usepackage{booktabs}
\usepackage{pifont}     
\usepackage{xcolor}
\usepackage{fontawesome}

\DeclareMathOperator*{\argmax}{arg\,max}
\definecolor{myyellow}{rgb}{0.788, 0.62, 0.063}

\makeatletter
\renewcommand\section[1]{%
  \par%
  \refstepcounter{section}%
  \textbf{#1---}%
  \hspace{0.5em}%
}
\makeatother
\begin{document}

\title{Improved belief propagation is sufficient for real-time decoding of quantum memory}
\author{Tristan Müller}
\email[Contact: ]{tristan.mueller@ibm.com}
\affiliation{IBM Quantum}
\author{Thomas Alexander}
\affiliation{IBM Quantum}
\author{Michael E. Beverland}
\affiliation{IBM Quantum}
\author{Markus~Bühler}
\affiliation{IBM Quantum}
\author{Blake R. Johnson}
\affiliation{IBM Quantum}
\author{Thilo Maurer}
\affiliation{IBM Quantum}
\author{Drew Vandeth}
\affiliation{IBM Quantum}

\date{\today}

\begin{abstract}
We introduce a new heuristic decoder, Relay-BP,  targeting real-time quantum circuit decoding for large-scale quantum computers. Relay-BP achieves high accuracy across circuit-noise decoding problems: significantly outperforming BP+OSD+CS-10 for bivariate-bicycle codes and comparable to min-weight-matching for surface codes. As a lightweight message-passing decoder, Relay-BP is inherently parallel, enabling rapid low-footprint decoding with FPGA or ASIC real-time implementations, similar to standard BP. 
A core aspect of our decoder is its enhancement of the standard BP algorithm by incorporating disordered memory strengths.  This dampens oscillations and breaks symmetries that trap traditional BP algorithms. By dynamically adjusting memory strengths in a relay approach, Relay-BP can consecutively encounter multiple valid corrections to improve decoding accuracy. We observe that a problem-dependent 
distribution of memory strengths that includes negative values is indispensable for good performance.
\end{abstract}

\maketitle

Quantum Error Correcting (QEC) codes protect information from noise to enable a fault-tolerant quantum computation.
Quantum Low Density Parity Check (qLDPC) codes, such as Bivariate Bicycle (BB) codes~\cite{kovalev2013quantum,bravyiEtAl2024}, are compelling candidates for fault-tolerant quantum computing~\cite{TremblayEtAl2022, bravyiEtAl2024, cross2024improvedqldpcsurgerylogical, williamson2024lowoverheadfaulttolerantquantumcomputation, he2025extractorsqldpcarchitecturesefficient}, offering lower qubit overhead than surface codes~\cite{kitaev_surface, KITAEV20032, bravyi1998quantumcodeslatticeboundary, PhysRevA.80.052312}.
For these codes to be practical, their circuit noise decoders must be fast enough to prevent backlog~\cite{Terhal_2015}, produce sufficiently low logical error rates, and be cost-effective.

We posit that for superconducting qubits with microsecond-scale QEC cycle times~\cite{Battistel_2023}, decoders must be implemented in Field-Programmable Gate Arrays (FPGAs) or potentially even in Application Specific Integrated Circuits (ASICs)~\footnote{ASICs enable similar parallel algorithm structures to FPGAs but tend to run faster at the price of being considerably more expensive and difficult to develop.}.
This introduces constraints for practical implementation: avoiding expensive matrix operations, eliminating global decisions during run-time, and preventing dynamic resizing of decoding submatrices.
Many CPU-efficient decoders are rendered less efficient on FPGAs or require significant adaptation.

Considerable effort has been dedicated to surface code decoding~\cite{Battistel_2023}, with leading approaches including Matching~\cite{kitaev_surface, Higgott2025sparseblossom, WuEtAl2023, Bausch_2024} and Union-Find~\cite{Delfosse2021almostlineartime}.
Matching involves irregular memory access and dynamic data structures and, to our knowledge, has no complete FPGA implementation~\cite{vittal2023astrea,wu2025micro}. 
Union-Find is slightly less accurate but involves simpler, local operations and has been implemented in FPGAs~\cite{ziad2024localclusteringdecoderfast,liyanage2023scalable}.
However, neither Matching nor Union-Find is flexible enough to decode circuits for general qLDPC codes as they rely on errors to trigger syndrome pairs, a feature specific to surface codes. 

Hence, we require that a real-time qLDPC decoder is:
\begin{itemize}[noitemsep]
     \item[(i)] \textit{Flexible}: Decodes a wide range of qLDPC circuits. 
     \item[(ii)] \textit{Accurate}: Has the ability to achieve the low logical error rates required by logical computations.
     \item[(iii)] \textit{Compact}: Has a small footprint on an FPGA.
    \item[(iv)] \textit{Fast}: Avoids the backlog problem by processing syndromes at their production rate.
    
\end{itemize}  

Belief Propagation (BP) is a flexible algorithm that decodes many classical LDPC codes well~\cite{RichardsonEtAl2008} and can also be applied to any qLDPC code.
As a message-passing algorithm, BP is ideal for compact and fast FPGA implementation due to its distributed memory and parallel processing.
However, BP often fails to converge for qLDPC codes when error beliefs oscillate due to loops and symmetric trapping sets caused by stabilizers (absent in classical LDPC codes)~\cite{iterativedecoding_poulin, trappingsets_raveendran} resulting in poor logical error rates as noted in Table~\ref{tbl:decoder-comparison}.

\begin{table}[h]
    \newcommand{\cmark}{\textcolor{green!70!black}{\ding{51}}}      
    \newcommand{\xmark}{\textcolor{red!80!black}{\ding{55}}}        
    \newcommand{\qmark}{\textcolor{myyellow!80!black}{\Large\textbf{?}}} 
    \newcommand{\circmark}{\textcolor{myyellow!80!black}{\faCircleO}} 

    \setlength{\tabcolsep}{3pt}
    \begin{tabular}{lcccc}
        \toprule
        \textbf{Decoder type} & \textbf{Flexible} & \textbf{Accurate}  & \textbf{Compact} & \textbf{Fast}     \\
        \midrule
        Matching & \xmark & \cmark & \qmark & \cmark  \\
        Union-Find & \xmark & \cmark & \cmark & \cmark \\
        BP/Mem-BP      & \cmark & \xmark & \cmark & \cmark  \\
        BP--OSD  & \cmark & \circmark & \xmark & \xmark  \\
        Clustering-based & \cmark & \circmark & \qmark & \circmark \\
        Search-based & \cmark & \cmark & \qmark & \circmark \\
        Relay--BP & \cmark & \cmark & \cmark & \cmark  \\
        \bottomrule
    \end{tabular}
    \caption{\label{tbl:decoder-comparison} 
    Selected decoder types given our requirements: flexible (applicable to any qLDPC code), accurate, compact (FPGA implementation), and fast.
    A circle is between a cross and a tick, and a question mark requires further research.
    }
\end{table}

Decoding qLDPC codes at all, even for benchmarking, was a significant challenge until the application of Ordered Statistics Decoding (OSD) by Panteleev and Kalachev~\cite{Panteleev_2021}.
They used BP+OSD decoding, where OSD follows the unconverged BP (indicated by `+'), utilizing partial decoding information from BP and Gaussian elimination to find a valid correction.
Improvements and an open-source implementation by Roffe et al.~\cite{decoding_roffe} established BP+OSD as the gold standard for qLDPC decoding.
Unfortunately, the enormous resources estimated for an FPGA implementation appear to render it infeasible for real-time decoding~\cite{syndrome_min_sum_arizona}.

Even on a CPU, OSD can be slow, prompting the development of faster follow-ups for BP. Ambiguity Clustering (AC)~\cite{ambiguity_clustering_wolanski} and Localized Statistics Decoding (LSD)~\cite{localized_statistics_decoding_hillmann} are examples of relatively fast clustering-based decoders proposed for this purpose, which are similar to Union-Find generalized to qLDPC codes~\cite{delfosse2022toward}.
Unfortunately, while Union-Find has a compact FPGA implementation for surface codes, its qLDPC generalizations require expensive matrix operations, albeit for smaller matrices than BP+OSD. Without further research, it is therefore unclear if clustering-based decoders can be implemented compactly on FPGAs for practical qLDPC decoding.
Recently, new search-based decoders~\cite{ott2025decisiontreedecodersgeneralquantum,beni2025tesseractsearchbaseddecoderquantum} and decimated-BP~\cite{gdg_gong} have emerged which are not only faster than BP-OSD, but also achieve significantly lower logical error rates.
However, these approaches also pose FPGA challenges due to their complex data structures.
Other attempts aimed at circumventing costly follow-ups for BP were unable to match these lower error rates \cite{koutsioumpas2025} or so far only tested with simplified noise models ~\cite{enhanced_message_passing_chytas,enhanced_min_sum_chytas}.

\begin{figure}[h!]
  \centering
  \includegraphics[width=1.0\linewidth]{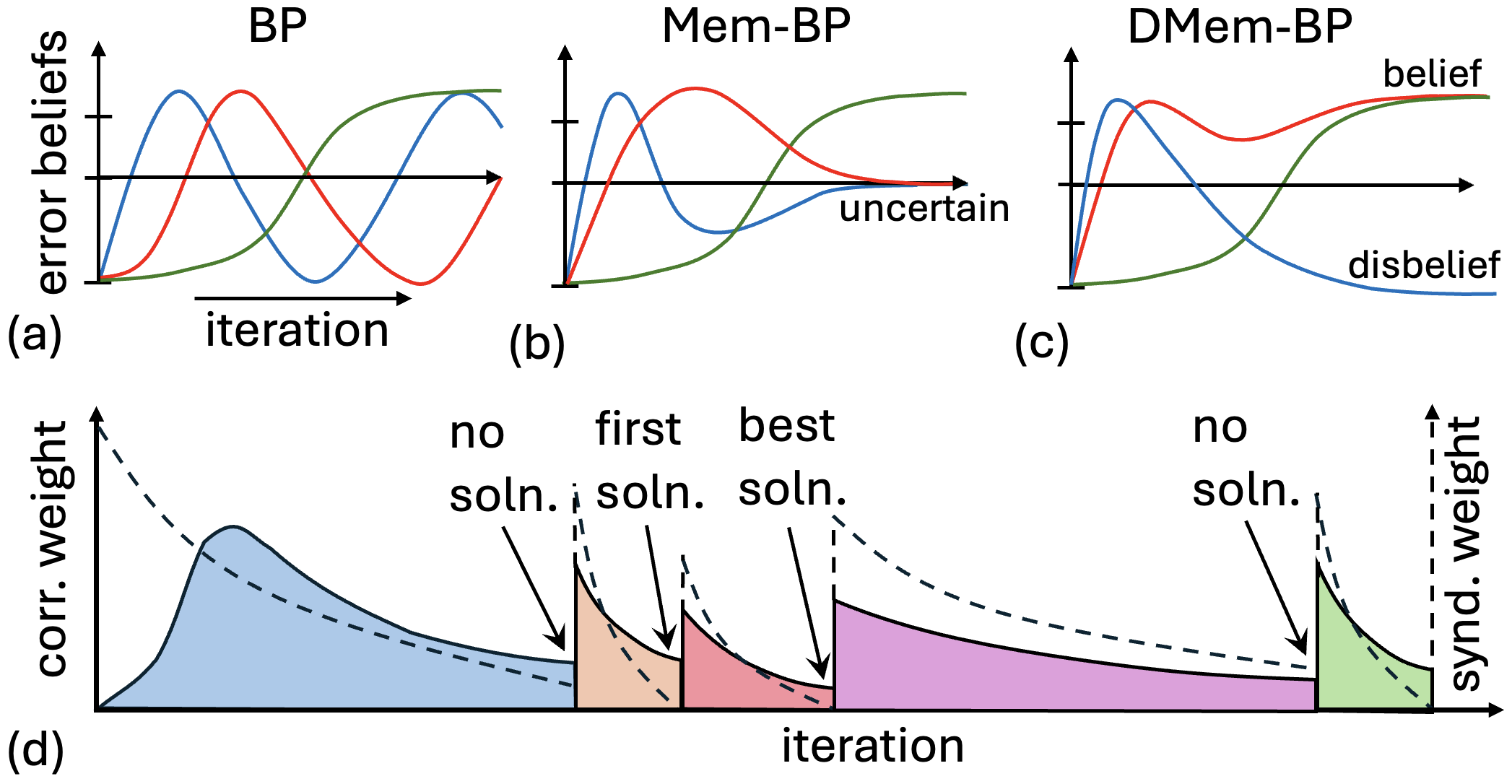}
  \caption{
  (a) In BP, the belief that each error occurred is updated over each iteration. 
  However, some beliefs can oscillate (red, blue) instead of converging (green). 
  (b) A memory term can dampen oscillations, but symmetric trapping sets may lead to convergence to uncertain beliefs (red, blue). 
  (c) Disordered memory strengths can break symmetries, leading to decisive beliefs forming a valid solution (i.e. the syndrome is canceled).
  (d) Relay-BP chains together different DMem-BP runs, which can further aid convergence and provide multiple valid solutions without restarting. 
  Solid lines indicate the weight of the proposed correction while dashed lines indicate the syndrome weight after the proposed correction.
  }
  \label{fig:relay_illustration}
\end{figure}

In this paper we introduce a new flexible qLDPC decoder called Relay-BP (``\textbf{R}elay-\textbf{e}nsembling with \textbf{l}ocally-\textbf{a}veraged memor\textbf{y}''), which satisfies all the criteria in Table~\ref{tbl:decoder-comparison}.
Relay-BP is based on modifications to BP that retain its message-passing structure thereby ensuring a compact and fast FPGA implementation. 
We first build on prior work that introduced memory terms to dampen oscillations in BP~\cite{loopy_bp_murphy,Nachmani2018,Kuo_2022,ewa_chen}, adopting a $\mathrm{GF}(2)$ version of EWAInit-BP~\cite{ewa_chen}, which we refer to as Mem-BP. Observing that varying memory strengths across node types improves convergence, we generalize this into DMem-BP using highly disordered memory strength distributions.  Secondly, a relay ensembling technique is incorporated in which successive DMem-BP runs are initialized with the previous run’s final marginals. Finally, to help escape from trapping sets, we allow the memory strengths to become negative. We illustrate these features in Figure~\ref{fig:relay_illustration}, and provide more explicit definitions and analysis of Relay-BP in the rest of the text.

\section{Decoding}
A decoding problem can be formally defined~\footnote{See the appendix for details or
reference~\cite{ott2025decisiontreedecodersgeneralquantum}} by a check matrix $\mathbf{H} \in \mathbb{F}_2^{M \times N}$, an action matrix $\mathbf{A} \in \mathbb{F}_2^{K \times N}$ and a probability vector $\mathbf{p} = (p_0, \dots, p_{N-1})$.
We assume each error location $j \in \{0, \dots, N-1 \} = [N]$ independently experiences an error with probability $p_j$. Let $\mathbf{e} \in \mathbb{F}_2^N$ be the  bit-string representing the unknown compound error and let $\bm{\sigma} = \mathbf{H} \mathbf{e} \in \mathbb{F}_2^M$ be the observed syndrome. 
Our task is to infer a correction $\mathbf{\hat{e}}$ using BP such that $\mathbf{H}\mathbf{\hat{e}}=\bm{\sigma}$ and $\mathbf{A\hat{e}} = \mathbf{A e}$. 
Different choices of $\mathbf{H}$, $\mathbf{A}$, and $\mathbf{p}$ define decoding problems across a wide range of settings: from classical codes to separate X and Z decoding for quantum CSS codes~\cite{steane1996multiple,calderbank1996good}, and further to fault-tolerant quantum circuits under correlated circuit noise. 
Let $G$ be the associated decoding graph where error nodes correspond to columns of $\mathbf{H}$ and check nodes correspond to rows of $\mathbf{H}$. We denote the neighbors of node $w$ in $G$ as $\mathcal{N}(w)$. 
A family of decoding problems is said to be qLDPC if the degree of $G$ does not grow with the problem size (we assume degree at most ten is desirable for practical message passing algorithms). 

\section{DMem-BP}
Relay-BP chains modified instances of Mem-BP, referred to as Disordered Memory Belief Propagation (DMem-BP).
We describe DMem-BP and clarify its relation to Mem-BP and standard BP~\footnote{BP refers to MinSum-BP as defined in the Factor Graphs chapter of Ref.~\cite{RichardsonEtAl2008}.}.

DMem-BP operates by iteratively sending real-valued messages along edges of the decoding graph $G$. Messages between check node $i$ and error node $j$ are denoted $\mu_{i\rightarrow j}$ and $\nu_{j\rightarrow i}$, each passing a local log-likelihood belief of whether an error $j$ occurred (negative) or not (positive). 
We specify the algorithm with its message update rules, initialization and stopping conditions. 

\emph{Messages:} The check-to-error messages at $t$ are:
\begin{equation}
\mu_{i \rightarrow j}(t) =   \kappa_{i,j}(t)\, (-1)^{\sigma_i} 
\min_{j' \in \mathcal{N}(i) \setminus \{ j \}}\left|
\nu_{j' \rightarrow i}(t-1)
\right|,
\label{eq:cntovn}
\end{equation}
\noindent where $\displaystyle \kappa_{i,j}(t) =  \mathrm{sgn}\bigg\{\prod_{j' \in \mathcal{N}(i) \setminus \{ j \}} 
\nu_{j' \rightarrow i}(t-1)\biggr\}$.\\
\noindent The error-to-check messages are then computed via:
\begin{equation}
    \nu_{j \rightarrow i}(t) =  {\displaystyle
    \Lambda_j(t) + \sum\limits_{i' \in \mathcal{N}(j) \setminus \{i\}} \mu_{i' \rightarrow j}(t).}
    \label{eq:vntocn}
\end{equation}

\noindent Conceptually, the message $\mu_{i \to j}(t)$ represents check $i$'s belief that error $j$ occurred, based on the check's received messages from its other neighboring error nodes. The sign is set by $\kappa_{i,j}(t)\, (-1)^{\sigma_i}$ to ensure consistency with syndrome $\sigma_i$, while the magnitude is set by the least confident (smallest magnitude) message that contributed to $\kappa_{i,j}(t)$. The message $\nu_{j \rightarrow i}(t)$ represents error node $j$'s belief an error occurred there, based on the sum of the messages $j$ received from other check nodes, shifted by a bias term $\Lambda_j(t)$ -- which we will discuss shortly.

\emph{Initialization and stopping:}
We initialize by setting the initial beliefs and biases as the log-likelihoods of the error priors $\Lambda_j(0)=\nu_{j\rightarrow i}(0) =\log\frac{1-p_j}{p_j}$ and set the initial marginals from user input $M_j(0)=M_j$. 
After each iteration of message passing, we compute the new marginal $M_j(t)$ and the hard decision $\hat{e}_j(t)$ for each error node $j$:
\begin{eqnarray}
    M_j(t) &=& {\displaystyle \Lambda_j(t) + \sum_{i'\in\mathcal{N}(j)}\mu_{i'\rightarrow j}(t),}
    \label{eq:posteriors} \\
       \hat{e}_j(t) &=&  {\displaystyle \text{HD} \bigl( M_j(t) \bigr), ~~ \text{for}~\text{HD}(x) = \tfrac{1}{2}\bigr(1-\text{sgn}(x)\bigr).} \nonumber
\end{eqnarray}

If the obtained error estimate $\hat{e}_j(t)$ satisfies the parity-check equation
$\mathbf{H}\bm{\hat{e}}(t)=\bm{\sigma}$, the algorithm is considered to have converged and the error vector $\hat{e}_j(t)$ is returned. 
Otherwise we move on to the next iteration until a maximum number of iterations $t = T$ has been reached, in which case DMem-BP is deemed unsuccessful.

\textit{Bias term:}
Under standard BP, the bias factors are fixed, and the initial marginals are chosen as $M_j=\Lambda_j(0)$, resulting in $\Lambda_j(t)=\Lambda_j(0)$ for all $t$. In DMem-BP different initial marginals are allowed and the biases are updated via the equation
\begin{equation}
    \Lambda_j(t)  = (1-\gamma_j) \Lambda_j(0) + \gamma_j M_j(t-1).
    \label{eq:relay_update}
\end{equation}
We use $\bm{\Gamma} = \{\gamma_j\}_{j\in[N]}$ to denote the real-numbers specifying the memory strength for each error node $j$.

DMem-BP therefore consists of a flexible family of decoders parameterized by the choice of memory strengths $\bm{\Gamma}$.
The special cases of all $\gamma_j$ values either zero or a constant between zero and one recover the standard BP and Mem-BP algorithms, which we compare against later.

\section{Relay-BP-$S$}The DMem-BP algorithm can be executed in parallel with initial marginals and varying memory strengths; however, its performance is typically improved when these instances are sequentially connected into a relay ensemble known as Relay-BP. Each DMem-BP instance will be called a \textit{leg} of the relay. To describe the relay ensemble properly we require some initial setup. For a given decoding problem $(\mathbf{H}, \mathbf{A}, \mathbf{p})$ the Relay-BP-$S$ algorithm is fully specified by the: 
\begin{itemize}[noitemsep]
    \item number of solutions sought $S$,
    \item maximum number of relay legs $R$, 
    \item maximum number of iterations for each leg $T_r$,
    \item memory strengths for each leg $\bm{\Gamma}_r = \{\gamma_j(r)\}_{j\in[N]}$,
\end{itemize}
\noindent 
for $r \in [R]$. 
The first leg of Relay-BP-$S$ applies DMem-BP initialized with marginals and memory strengths $M_j(0) = \log\frac{1-p_j}{p_j}$ and $\bm{\Gamma}_0= \{\gamma_j(0)\}_{j\in[N]}$. 
The $r^{th}$ leg's marginals are initialized with the $(r-1)^{th}$ leg's final marginals, thereby passing information forward through the relay. Each leg stops upon finding a solution or reaching the iteration limit $T_r$. The algorithm stops when $R$ legs have executed or $S$ solutions have been found, and returns the lowest-weight solution found among all legs, where solution $\mathbf{\hat{e}}$ has weight $w(\mathbf{\hat{e}}) = \sum_j  \hat{e}_j \log\frac{1-p_j}{p_j}$. Pseudocode for Relay-BP is included in Appendix, and the implementation source code is available on GitHub~\footnote{\url{ https://github.com/trmue/relay}}.

\begin{figure*}[t]
  \centering
  \includegraphics[width=0.98\linewidth]{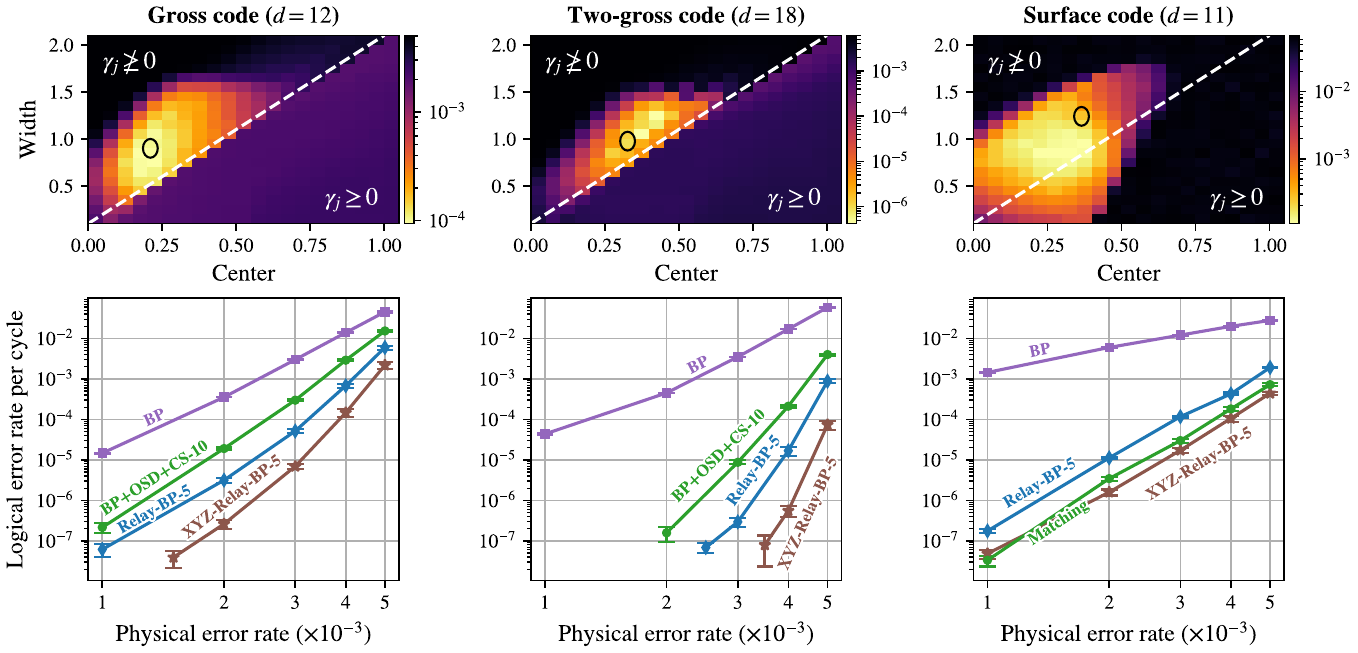}
  \caption{
    Circuit-noise decoding examples.
    (Top) Relay-BP-1 logical error rate heatmaps at $p = 3 \times 10^{-3}$ vs. memory strength intervals.
    Circles mark intervals used for decoding: $[-0.24, 0.66]$ (gross), $[-0.161, 0.815]$ (two-gross), $[-0.254, 0.985]$ (surface). The dashed line indicates a threshold above which negative $\gamma_j$ are present.
    (Bottom) Relay-BP outperforms BP+OSD+CS-10 on the gross and two-gross codes, and performs comparably to Matching on the surface code.
    Top panels share their width scale but have separate logical error rate heat maps; bottom panels share a logical error rate scale.
  }
  \label{fig:relay_panel}
\end{figure*}

\section{Decoding examples}
In Figure.~\ref{fig:relay_panel},
we evaluate Relay-BP under circuit-level noise on two Bivariate Bicycle~\cite{kovalev2013quantum} (BB) codes—the $[[144,12,12]]$ \textit{gross} code and the $[[288,12,18]]$ \textit{two-gross} code (defined and implemented using circuits from Ref.~\cite{bravyiEtAl2024})—and the distance-11 rotated surface code~\cite{bravyi1998quantumcodeslatticeboundary}, implemented using circuits from Ref.~\cite{Gidney_2021}.
These examples are all CSS-type. 
Following standard practice, each simulation uses $d$ noisy QEC cycles with strength-$p$ circuit noise followed by one perfect cycle; details, including the construction of the decoding objects $\mathbf{H}$, $\mathbf{A}$, and $\mathbf{p}$, are provided in the appendix.

We consider two decoding strategies. 
\emph{XYZ-decoding} directly computes a correction $\mathbf{\hat{e}}$ from $\mathbf{H}$ and $\bm{\sigma}$. \emph{XZ-decoding} decomposes $\bm{\sigma}$ into $\bm{\sigma}_X$ and $\bm{\sigma}_Z$, which are independently decoded using the derived check matrices $\mathbf{H}_X$ and $\mathbf{H}_Z$ to obtain $\mathbf{\hat{e}}_X$ and $\mathbf{\hat{e}}_Z$. 
These partial corrections are then combined to infer $\mathbf{\hat{e}}$. 
XZ-decoding simplifies decoding by using smaller objects but may degrade performance by modeling X and Z errors as independent. 
All three decoding examples are CSS-type; we use XZ-decoding throughout unless explicitly stated otherwise.
The XZ-decoding matrices have approximate dimensions 1k$\times$9k, 2k$\times$26k and 1k$\times$7k for gross, two-gross and surface code examples respectively, with 2k$\times$71k, 5k$\times$213k and 1k$\times$24k for the XYZ-decoding matrices.

In all Relay-BP simulations, we set the max iterations to $T_r = 60$ for each leg, except the first, which uses $T_0 = 80$ due to slower initial convergence. 
The first leg uses a uniform memory strength of $\gamma = 0.125$ for the gross and two-gross codes, and $\gamma = 0.35$ for the surface code, based on preliminary sweep results.
In all subsequent legs, each $\gamma_j$ is drawn independently and uniformly from a common range $[\gamma_\text{center} - \gamma_\text{width}/2, \gamma_\text{center} + \gamma_\text{width}/2]$. 

\section{Memory strength selection}
Relay-BP’s accuracy can be improved significantly by appropriate  problem-dependent choice of memory strengths. 
Figure~\ref{fig:relay_panel} shows heatmaps for each of the three decoding problems at $p = 3 \times 10^{-3}$, reporting logical error rates of Relay-BP-1 ($R = 301$) with memory strengths from intervals with various values of $\gamma_\text{center}$ and $\gamma_\text{width}$.
Each tile uses the smaller of 500,000 samples or a number achieving $\leq 10\%$ relative accuracy.

A striking feature in the heatmap for the gross and two-gross codes (and to a lesser extent for the surface code) is a performance hotspot above the diagonal, corresponding to intervals that include negative memory strengths.
Intervals marked with circles in the heatmaps were obtained via a gradient-free optimization and are used for all subsequent simulations.

\section{Flexible decoding}
Relay-BP achieves low logical error rates across all three decoding examples. Figure~\ref{fig:relay_panel} compares both XZ- and XYZ-decoding to standard benchmarks. 
For the BB codes, we compare to BP+OSD+CS-10 (via the LDPC package~\cite{Roffe_LDPC_Python_tools_2022} using 10,000 BP iterations, with a combination-sweep (CS) of 10); for the surface code, to Matching using Sparse Blossom in PyMatching~\cite{Higgott2025sparseblossom}. 
Standard BP is included for additional context (with a maximum of 10,000 iterations).
We set $R = 301$ for Relay-BP-1 but $R = 601$ for Relay-BP-5 to give it more opportunity to find multiple solutions.

For the gross code, both Relay-BP-5 and XYZ-Relay-BP-5 outperform BP+OSD+CS-10 across the full range.
At $p = 3 \times 10^{-3}$, they yield improvements of approximately one and two orders of magnitude, respectively.
For the two-gross code at the same $p$, we see even larger improvements for Relay-BP-5 and would predict even larger gains for XYZ-Relay-BP-5.
While we would not advocate Relay-BP for real-time surface code decoding due to the availability of other good options, its performance is encouraging. 
Relay-BP achieves logical error rates several orders of magnitude lower than standard BP as $p$ decreases. 
XYZ-Relay-BP-5 outperforms Relay-BP-5 and performs comparably to Matching~\footnote{PyMatching’s CPU implementation of Sparse Blossom is substantially faster than our CPU implementation of XYZ-Relay-BP-5.} although we note that iterative Matching variants that go beyond XZ-decoding can achieve better performance~\cite{paler2023pipelined,fowleroptimal}.

Finally, one might ask whether Relay-BP’s advantage arises solely from ensembling, rather than its relay structure. 
To test this, we compare two different ensembling methods, both performing XYZ-decoding with $R = 601$ on the gross code with $p=3\times 10^{-3}$. 
In the Relay-BP version, each leg is initialized with the output marginals of the previous leg as usual. 
In the independent ensembling variant, each leg instead restarts from the original priors.
We observe that the standard version achieves a logical error rate of $(7 \pm 1) \times 10^{-6}$ with an average of $330.8 \pm 0.5$ iterations, whereas independent ensembling yields $(1.4 \pm 0.2) \times 10^{-5}$ with $578 \pm 2$ iterations.
Relay ensembling therefore improves both convergence rate and solution quality.

\begin{figure}[t]
    \includegraphics[width=0.99\linewidth]{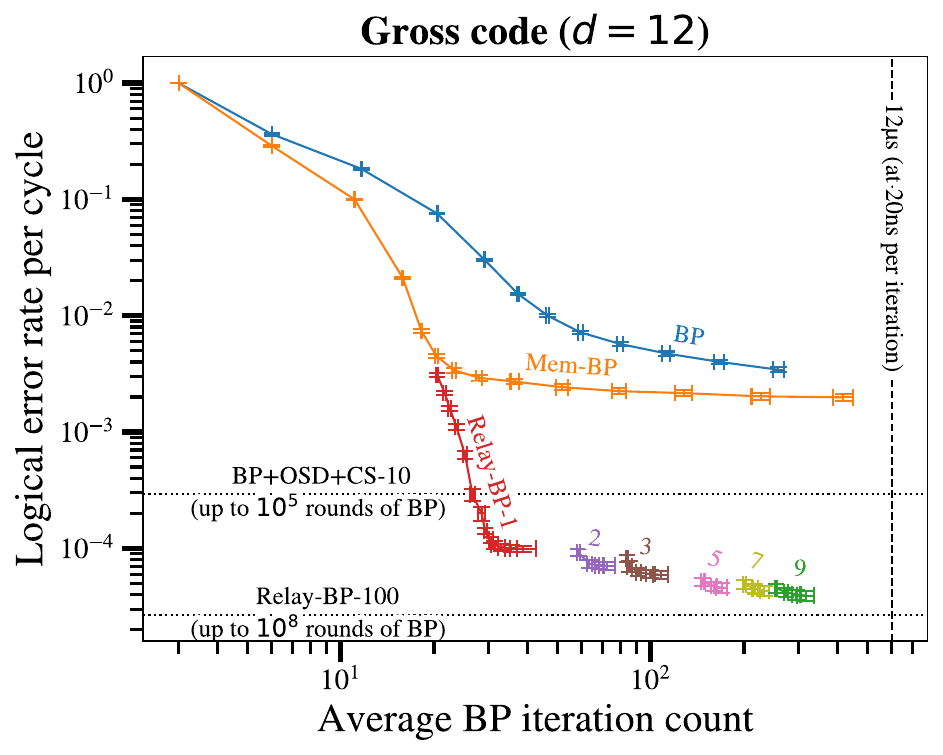}
    \caption{\label{fig:decoder_results}
    Logical error rate vs. average number of BP iterations at $p = 3 \times 10^{-3}$.
    Relay-BP-$S$ curves are generated by varying the maximum number of legs $R$; BP and Mem-BP curves by varying their maximum iteration count.
    Relay-BP achieves substantially lower error rates than other decoders within the estimated 600-iteration real-time budget.
    }
\end{figure}

\section{Real-time decoding}
We assess the real-time feasibility of Relay-BP using XZ-decoding of the gross code at $p = 3 \times 10^{-3}$.
This 12-cycle circuit at 1~$\mu$s per cycle can accommodate $\sim 600$ BP iterations to achieve real-time decoding without backlog. 
This estimate assumes a 20~ns iteration time, based on Ref.~\cite{syndrome_min_sum_arizona}, which reports 16~ns for a slightly simpler decoding problem using a fully parallel FPGA implementation with dedicated resources for each variable and check node. 

Figure~\ref{fig:decoder_results} shows the average number of BP iterations versus logical error rate for Relay-BP and other BP-based decoders. 
Mem-BP initially improves over BP but saturates near 50 iterations, offering only modest gains. 
In contrast, Relay-BP-1 achieves error rates orders of magnitude lower than BP and Mem-BP within 30 iterations, and outperforms a high-resource implementation of BP+OSD+CS-10 by a factor of three.
Increasing $S$ improves performance: at 300 iterations (well within the 600-iteration budget), Relay-BP-9 achieves a twofold improvement over Relay-BP-1 and nearly matches the result of a high-resource Relay-BP-100 run, indicating diminishing returns beyond $S = 9$ in this scenario.
These results demonstrate that Relay-BP meets real-time decoding requirements while maintaining high accuracy.

In Figure~\ref{fig:decoder_results_XYZ} in the appendix, we observe that with 80 iterations, Relay-BP-1 achieves an order of magnitude lower logical error rate for XYZ-decoding than for XZ-decoding on the same problem.
While we expect the FPGA iteration time only grows moderately with decoding matrix dimensions, we do expect XYZ-decoding ($\sim$2k$\times$71k) to be somewhat slower than XZ-decoding ($\sim$1k$\times$9k), reducing the effective iteration budget below the 600 estimated for XZ.
For larger codes, the increased problem size will further raise iteration time and may exceed a single FPGA’s capacity if using the fully parallel implementation strategy of Ref.~\cite{syndrome_min_sum_arizona}.
This work focuses on quantum memory, where avoiding the backlog problem requires a low average-case decoding time.
Logical operations—especially those involving multiple code blocks—not only increase problem size but may also require decoding to complete before the next logical operation is selected, such that the tail of the decoding time distribution is also crucial~\cite{BicycleArchitecturePaper}.

\section{Discussions and Outlook}
In summary we have introduced a new heuristic decoder, called Relay-BP, which chains together multiple BP runs, each with different disordered memory strengths.
The Relay-BP decoder can achieve orders of magnitude better logical error rates than BP+OSD+CS-10 when decoding LDPC codes and for the rotated surface code Relay-BP improves the error rates of standard BP by multiple orders of magnitude, achieving comparable error rates to matching-based decoders. Lastly, all evidence suggests that Relay-BP is fast enough for real-time decoding on an FPGA while achieving very low logical error rates.

More work needs to be done and there are many unanswered questions: Why are negative memory strengths so important? How well does Relay-BP perform beyond memory experiments? Are there better ways to choose memory strengths? How well can a full FPGA implementation perform? We hope to answer these questions and many more in future work.

\section{Acknowledgments}
The authors would like to thank 
Lev Bishop,
Paul Bye,
Andrew Cross, 
Jay Gambetta,
Frank Haverkamp,
Tomas Jochym-O'Connor, 
Anirudh Krishna, 
Michael Kröner 
and Ted Yoder.

\bibliography{biblio.bib}

\appendix*
\begin{center}\textbf{Appendix}\end{center}

\section{Decoding objects for logical memory circuits} 
\begin{figure}[h!]
    \includegraphics[width=0.9\linewidth]{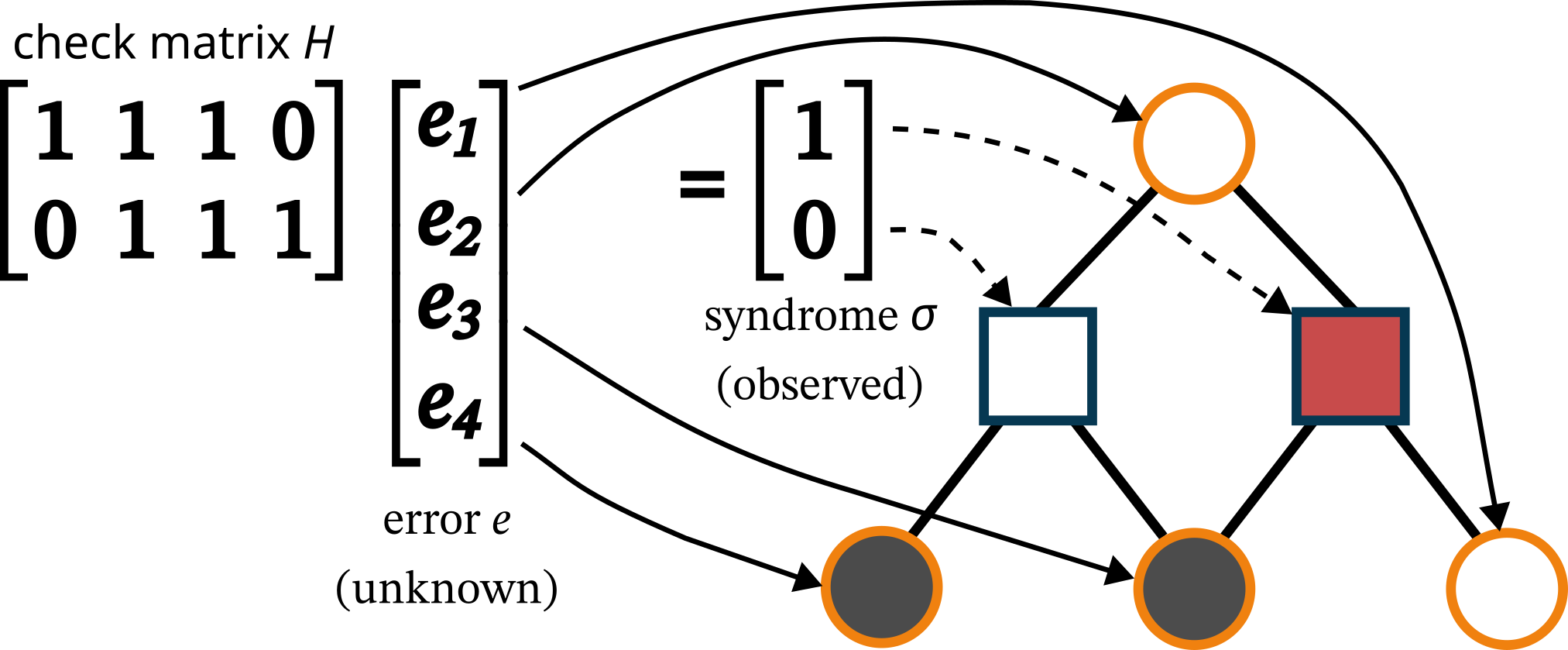}
    \caption{\label{fig:decoding-basics} The decoding graph visually represents the check matrix: circular ($\ocircle$) \emph{error nodes}  correspond to columns of $\mathbf{H}$, and square ($\square$) \emph{check nodes}  to its rows. 
    Filled nodes denote value 1 and unfilled nodes value 0. 
    The decoder relies solely on the syndrome (i.e., the values of the check nodes) to infer a candidate error. 
    Check node $i$ has syndrome $\sigma_i = 1$ if it touches an odd number of error nodes with value one.}
\end{figure}

We follow the general decoding framework presented in~\cite{ott2025decisiontreedecodersgeneralquantum} to produce the decoding objects $\mathbf{H}$, $\mathbf{A}$ and $\mathbf{p}$ which specify the decoding problem. 
Figure~\ref{fig:decoding-basics} illustrates the corresponding decoding graph. 

Our decoding examples are logical memory circuits based on an $[[n,k,d]]$ stabilizer code with stabilizer group $S$.
The overall circuit is built by repeatedly applying a small circuit (called a QEC cycle) that measures each of a fixed set of $m$ stabilizer generators.
In our simulations, we take the standard approach that the QEC cycle is repeated $C=d$ times with noise present, followed by an additional final cycle which occurs error-free (a condition that can be relaxed with windowed decoding). 
This produces $M = C \times m$ binary outcomes known as circuit checks or ‘detectors’ \cite{Gidney_2021}, whose values are the parity of outcomes across successive rounds (0 if satisfied, 1 otherwise). 

We assume a linear circuit-level noise model in which each quantum operation fails independently, with errors sampled from a discrete set of error modes. 
Each single-qubit unitary is followed by a Pauli $X$, $Y$, or $Z$ error, each with probability $p/3$, and each two-qubit unitary by one of the 15 non-identity two-qubit Paulis with probability $p/15$. 
State preparations and measurements fail with probability $p$, modeled respectively by orthogonal state preparation or measurement outcome flipping. 
We refer to each such error mode as an error, which corresponds to a column in $\mathbf{H}$ and $\mathbf{A}$, with its probability encoded in the associated entry of the vector $\mathbf{p}$.

Each error results in a subset of detector outcomes flipping, and induces a residual Pauli at the end of the circuit. 
These effects define the matrices $\mathbf{H}$ and $\mathbf{A}$ as follows.
$\mathbf{H}_{i,j} = 1$ iff the $j$th error flips the $i$th detector.
$\mathbf{A}_{l,j} = 1$ iff the residual Pauli $E_j$ of the  $j$th error anti-commutes with the $l$th stabilizer generator $S_l$.

Once the XYZ-decoding objects $\mathbf{H}$, $\mathbf{A}$ and $\mathbf{p}$ have been obtained, we can further obtain the XZ-decoding objects $(\mathbf{H}_X,\mathbf{A}_X,\mathbf{p}_X)$ and $(\mathbf{H}_Z,\mathbf{A}_Z,\mathbf{p}_Z)$ as follows.
First, note that every row in $\mathbf{H}$ is either X or Z type, as it corresponds to a detector obtained from the parity of a pair of X-type or Z-type stabilizer generators.
We construct $\mathbf{H}_X$ by simply extracting rows from $\mathbf{H}$ corresponding to Z-type detectors.
Similarly, we construct $\mathbf{A}_X$ by simply extracting rows from $\mathbf{A}$ corresponding to Z-type stabilizer generators.
$\mathbf{p}_X$ is identical to $\mathbf{p}$.
A similar process is used to construct $(\mathbf{H}_Z,\mathbf{A}_Z,\mathbf{p}_Z)$.

Lastly, we compress the decoding objects as follows. 
(Note that this step is applied to each set of decoding objects independently, and can lead to the XZ decoding objects having far fewer columns than the XYZ decoding objects.)
Start with $\mathbf{H}$ (or $\mathbf{H}_X$ or $\mathbf{H}_Z$). 
Start with $j=0$.
Let $\mathcal{J} = \{ j,j_2,\dots,j_R \}$ be the set of all column labels that have identical columns in $\mathbf{H}$ (i.e., $\mathbf{H}_{*,j'}$ is the same for all $j' \in \mathcal{J}$).
Note that provided the circuit distance is two or above (required for any error correction capability) the corresponding columns of the action matrix must be identical too, i.e., $\mathbf{A}_{*,j'}$ is the same for all $j' \in \mathcal{J}$.
Compute $p'_j$, the probability that an odd number of errors indexed by $\mathcal{J}$ occur. 
We update $\mathbf{H}$, $\mathbf{A}$ and $\mathbf{p}$ by removing all columns from each which are indexed by $\mathcal{J}\setminus \{ j \}$, and set the $j$th entry of $\mathbf{p}$ to  $p'_j$. 
Move on to $j+1$ and repeat this process until all columns of $\mathbf{H}$ are unique.

\section{Decoding}
Let $\mathbf{p}\in(0,1/2)^N$ be the \emph{probability vector} where $p_j$ is the probability that error $j$ occurs. The discrete noise is then represented as an error vector $\mathbf{e}\in \mathbb{F}_2^N$, where we make the assumption that each bit is drawn independently according to a probability vector $\mathbf{p}$, such that:
\begin{eqnarray*}
\text{Pr}_{\mathbf{p}}(\mathbf{e}) =  \prod_{j =0}^{N-1} (1-p_j) \left(\frac{p_j}{1-p_j}\right)^{e_j}.
\end{eqnarray*}
\noindent We follow the standard approach by not trying to estimate the maximum-likelihood error and instead estimate the bit-wise most likely error:
$
\mathbf{\hat{e}}_{\text{bw}} =
\left[
\argmax_{x_i \in\mathbb{F}_2}\,
\text{Pr}_{\mathbf{p}}(x_i|\boldsymbol{\sigma})
\right]_{i=0}^{N-1}.
$
Here $\text{Pr}_{\mathbf{p}}(x_i|\boldsymbol{\sigma})$ are the marginals of $\text{Pr}_{\mathbf{p}}(\mathbf{x}|\boldsymbol{\sigma})$ as defined by:
\[
\text{Pr}_{\mathbf{p}}(x_i|\boldsymbol{\sigma}) =
\sum_{\sim\{x_i\}} \text{Pr}_{\mathbf{p}}(x_0,x_1,\dots,x_{N-1}|\boldsymbol{\sigma}).
\]
These marginals will be calculated using the BP algorithm which in turn requires a suitable factorization of $\text{Pr}_{\mathbf{p}}(\mathbf{x}|\boldsymbol{\sigma})$. This can be done using our assumption of independence of errors from which we see that:
\begin{equation*}
    \text{Pr}_{\mathbf{p}}(\mathbf{x}|\boldsymbol{\sigma}) \propto
    \left(
        \prod_{j=0}^{N-1} \text{Pr}_{p_j}(x_i)
    \right)
    \left(
        \prod_{k=0}^{M-1} [\mathbf{H}_k\mathbf{x}=\sigma_k]
    \right), \label{eq:factorization}
\end{equation*}
\noindent where $\mathbf{H}_k$ is the $k$-th row of $\mathbf{H}$. Here we have used Iverson’s notation~\cite{Iverson1962}. That is $[P]$ is $1$ if $P$ is true and $[P]=0$ otherwise.

\section{XYZ-decoding real-time analysis}
\begin{figure}[t]
    \includegraphics[width=0.99\linewidth]{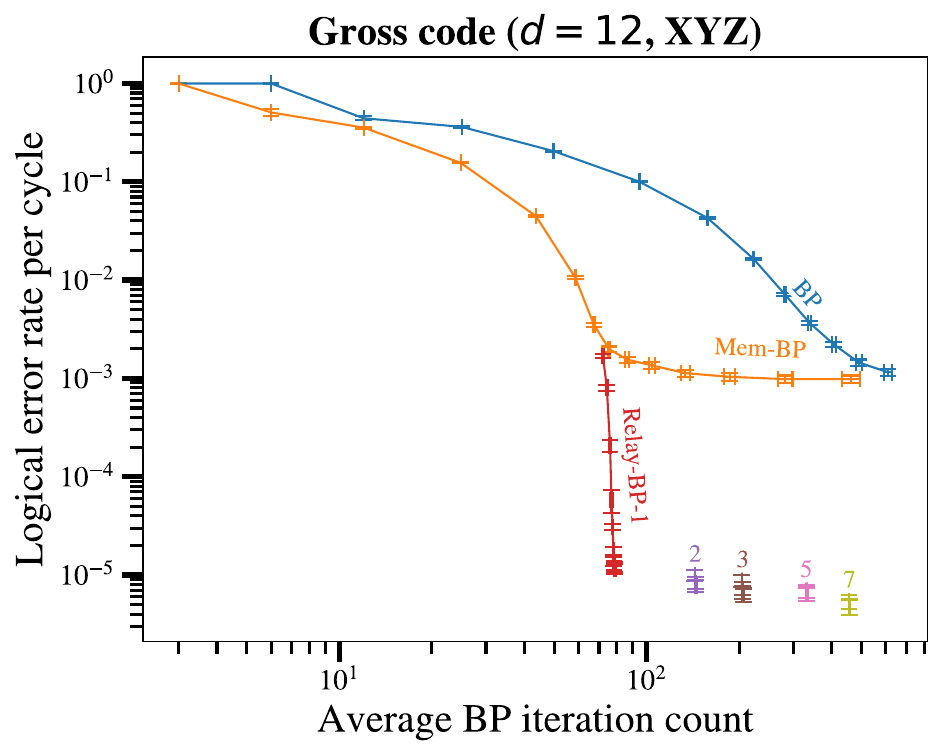}
    \caption{\label{fig:decoder_results_XYZ}
    Logical error rate vs. average number of BP iterations at $p = 3 \times 10^{-3}$, with XYZ-decoding.
    XYZ-Relay-BP-$S$ curves are generated by varying the maximum number of legs $R$; XYZ-BP and XYZ-Mem-BP curves by varying their maximum iteration count.
    }
\end{figure}

In Figure~\ref{fig:decoder_results_XYZ} we repeat the analysis presented in Figure~\ref{fig:decoder_results} in the main text, but using XYZ decoding instead of XZ decoding.

\FloatBarrier

\section{Pseudo-Code for Relay Algorithm}
For simplicity, the universal quantifier $\forall j$ is dropped in the following pseudo-code when the intention is clear. 

\begin{algorithm}[h]
    \SetKwInOut{Input}{Input}
    \SetKwInOut{Output}{Output}
    \Input{Parity-check matrix $\boldsymbol{\mathrm{H}}$, syndrome $\boldsymbol{\sigma}$, error probabilities $\boldsymbol{\mathrm{p}}$, number of solutions to be found $S$, maximum number of legs of the relay $R$, maximum number of iterations per leg $T_r$, and memory strengths for each leg $\{\boldsymbol{\gamma}(r)\}_{r\in[R]}$.}
    \Output{Solution found, Estimated error $\mathbf{\hat{e}}$}
    $\lambda_j, M_j \left( 0 \right) \leftarrow \log\frac{1-p_j}{p_j}$, $r \leftarrow 0,  s \leftarrow 0, \hat{\mathbf{e}} \leftarrow 	\varnothing, \omega_{\hat{\mathbf{e}}} \leftarrow \infty$\;
    \For{$r \leq R$}
      {
        \tcp{Run DMem-BP}
        $\Lambda_j(0) \leftarrow \nu_{j\rightarrow i}(0) \leftarrow \lambda_j$\;
        \For{$t \leq T_r$}
          {     
            $\Lambda_j(t) \leftarrow (1-\gamma_j(r)) \Lambda_j(0) + \gamma_j(r) M_j(t-1)$\;
            Compute $\mu_{i \rightarrow j} \left( t \right)$ \tcp*[f]{via Eq.~\eqref{eq:cntovn}}\;
            Compute $\nu_{j \rightarrow i} \left(t \right)$  \tcp*[f]{via Eq.~\eqref{eq:vntocn}}\;
            Compute $M_j \left(t \right)$  \tcp*[f]{via Eq.~\eqref{eq:posteriors}}\;
            $\hat{e}_j(t) \leftarrow \mathrm{HD} \bigl( M_j(t) \bigr)$\;
            \If{$\mathbf{H\hspace{0.0833em}\mathbf{\hat{e}}}(t) = \boldsymbol{\sigma}$}
            {
              \tcp*[h]{BP converged}\;
              $\omega_r \leftarrow w(\mathbf{\hat{e}}) = \sum_j  \hat{e}_j \lambda_j$\;
              $s \leftarrow s + 1$\;
              \If{$\omega_r < \omega_{\hat{\mathbf{e}}}$}
              {
                $\hat{\mathbf{e}} \leftarrow \hat{\mathbf{e}}(t)$\;
                $\omega_{\hat{\mathbf{e}}} \leftarrow \omega_r$\;
              }
              $\boldsymbol{\mathrm{break}}$; \tcp*[f]{Continue to next leg}
            }
            $t \leftarrow t + 1$\;
          }
        \If{$s = S$}
            {
              $\boldsymbol{\mathrm{break}}$; \tcp*[f]{Found enough solutions}
            }
        \tcp{Reuse final marginals for the next leg}
        $M_j(0) \leftarrow M_j \left( t \right)$\;
        $r \leftarrow r + 1$\;
      }
    $\boldsymbol{\mathrm{return}}$ $(s > 0)$, $\mathbf{\hat{e}}$;
     
    \caption{Relay-BP-$S$ decoder for quantum LDPC codes}
    \label{alg:Relay}
\end{algorithm}
\FloatBarrier

\end{document}